\documentclass{jpsj2}
\usepackage{graphicx}
\usepackage{subfigure}
\usepackage{mcite}

\title{
Effect of Fermi Surface Topology on Inter-Layer Magnetoresistance in
Layered Multiband Systems: Application to LaFeAsO$_{1-x}$F$_x$
}

\author{
Takao \textsc{Morinari}$^{1}$\thanks{morinari@yukawa.kyoto-u.ac.jp},
Hiroki \textsc{Nakamura}$^{2,3,4}$,
Masahiko \textsc{Machida}$^{2,3,4}$,
and Takami \textsc{Tohyama}$^{1,4}$
}

\inst{
$^{1}$Yukawa Institute for Theoretical Physics, Kyoto University, Kyoto 606-8502, Japan\\
$^{2}$CCSE, Japan Atomic Energy Agency, 6-9-3 Higashi-Ueno, Tokyo 110-0015, Japan\\
$^{3}$CREST(JST), 4-1-8 Honcho, Kawaguchi, Saitama 332-0012, Japan\\
$^{4}$JST, Transformative Research-Project on Iron Pnictides (TRIP), Chiyoda, Tokyo 102-0075, Japan\\
}

\abst{
In layered single band systems, 
the interlayer conductivity depends
on the orientation of the in-plane magnetic field
and takes maximum values when the magnetic field is perpendicular to 
flat regions of the Fermi surface.
Extending this known results to multi-band systems,
we propose an experiment 
to extract information about their Fermi surface topology.
We discuss application of the formula to a FeAs-based superconductor, 
LaFeAsO$_{1-x}$F$_x$.
We show that the magnetically ordered state in the parent compound
is clearly distinguished from the paramagnetic state
by the oscillation period in the interlayer conductivity.
We demonstrate that
evolution of the Fermi surface topology by changing the doping concentration
is reflected to the interlayer conductivity oscillation patterns.
}

\kword{
Fermi surface, interlayer magnetoresistance, Fe-pnictide
}

\newcommand{\refeq}[1]{(\ref{#1})}                                  
\newcommand{\be}{\begin{equation}}
\newcommand{\ee}{\end{equation}}
\newcommand{\bea}{\begin{eqnarray}}
\newcommand{\eea}{\end{eqnarray}}
\begin{document}
\maketitle

\section{Introduction}
Determination of the Fermi surface topology plays an important role
to make clear the basic electronic structure of a superconductor.
For layered compounds, angle-resolved photoemission spectroscopy (ARPES)
can be used to study the Fermi surface.
Although ARPES provides the map of the Fermi surface,
ARPES measurements are sensitive to surface conditions.
Contrary, magnetic quantum oscillations provide information
about the Fermi surface area perpendicular to the applied magnetic field.
The Fermi surface areas determined by magnetic quantum oscillations
are bulk electronic properties but this technique does not provide
information about the Fermi surface topology directly.
In this paper, we propose interlayer magnetoresistance experiments
to probe the bulk Fermi surface topology 
in layered multi-band systems
and its application
to the newly discovered FeAs-based layered superconductors.\cite{Kamihara08}
To be specific, we focus on LaFeAsO$_{1-x}$F$_x$
and discuss what we expect for the interlayer magnetoresistance
in the paramagnetic state and the magnetically ordered state.

%\subsection{Fermi surface study in FeAs-superconductors}
% First principle calculations
% Normal state and magnetic state
% magnetic state: Cruz2008
% ARPES
% de Haas-van Alphen 
% materials
The Fermi surfaces of the FeAs-based layered superconductors have been
studied theoretically and experimentally.
First principles calculations \cite{Singh08,Ishibashi08}
% other papers: Kuroki08[15-18]
suggest that the Fermi surface of LaFeAsO consists 
of hole Fermi surfaces around $\Gamma$ point and 
electron Fermi surfaces around $M$ point.
The Fermi surfaces measured by ARPES\cite{Zabolotnyy09,Lu08,Liu08,Ding08,Kondo08} 
are mostly consistent with this picture,
though the measured samples are limited.
%%%%%%%%%%%%%%%%%%%%%%%%%%%%%%%%%%%%%%%%%%%%%%%%%%%%%%%%%%%%%%%%%%%%%%%%%%%
% Revise 0729/09
%%%%%%%%%%%%%%%%%%%%%%%%%%%%%%%%%%%%%%%%%%%%%%%%%%%%%%%%%%%%%%%%%%%%%%%%%%%
%{\bf 
For example, a single hole pocket and a single electron pocket
were observed in NdFeAsO$_{0.9}$F$_{0.1}$.\cite{LiuKondo08,Kondo08},
while the other Fermi pockets predicted by the band calculations
were not observed.
%and LaFeAsO has not been measured yet.
The hole Fermi surfaces and the electron Fermi surfaces 
have instability of a magnetic order
%}
%%%%%%%%%%%%%%%%%%%%%%%%%%%%%%%%%%%%%%%%%%%%%%%%%%%%%%%%%%%%%%%%%%%%%%%%%%%
% end
%%%%%%%%%%%%%%%%%%%%%%%%%%%%%%%%%%%%%%%%%%%%%%%%%%%%%%%%%%%%%%%%%%%%%%%%%%%
associated with a nesting of $(\pi,\pi)$ wavevector.
The magnetically ordered state is a collinear antiferromagnetic
state as suggested from the neutron scattering experiments.\cite{Cruz08,Huang08}
Quantum oscillation measurements were carried out 
for magnetic states of a few Fe-based superconductors.\cite{Coldea08,Sugawara08,Sebastian08}
Possible folding of Brillouin zone are discussed 
so that the observed oscillation periods are reproduced.

%\subsection{Review of interlayer magnetoresistance}
Another experiment to probe the Fermi surface of layered compounds
is to measure angular magnetoresistance oscillations
%%%%%%%%%%%%%%%%%%%%%%%%%%%%%%%%%%%%%%%%%%%%%%%%%%%%%%%%%%%%%%%%%%%%%%%%%%%
% Revise 0729/09
%%%%%%%%%%%%%%%%%%%%%%%%%%%%%%%%%%%%%%%%%%%%%%%%%%%%%%%%%%%%%%%%%%%%%%%%%%%
%{\bf 
which was firstly discovered in the organic conductors
$\beta$-(BEDT-TTF)$_2$IBr$_2$\cite{Kartsovnik88}
and 
$\theta$-(BEDT-TTF)$_2$I$_3$.\cite{Kajita89}
The magnetoresistance oscillations in these systems
were well understood by a semiclassical analysis
assuming an energy dispersion
described by a two-dimensional cylinder with
a weak cosine warping.\cite{Yamaji89,Yagi90}
In this magnetoresistance phenomenon,
if the applied magnetic field 
lies in the plane,\cite{Lebed97,Schofield00,Dragulescu99}
the oscillation patterns are associated with
the Fermi surface topology.
%}
%%%%%%%%%%%%%%%%%%%%%%%%%%%%%%%%%%%%%%%%%%%%%%%%%%%%%%%%%%%%%%%%%%%%%%%%%%%
% End
%%%%%%%%%%%%%%%%%%%%%%%%%%%%%%%%%%%%%%%%%%%%%%%%%%%%%%%%%%%%%%%%%%%%%%%%%%%
The interlayer conductivity is dominated by
the regions of the Fermi surface where the Fermi velocity is
parallel to the magnetic field.
The theoretical calculation for Tl$_2$Ba$_2$CuO$_6$\cite{Dragulescu99}
is consistent with the experiment.\cite{Hussey96}
% Dragulescu99[5]

%\subsection{Purpose of this paper}
In this paper, we extend the formula for the single-band 
systems \cite{Lebed97,Schofield00,Dragulescu99} to multi-band systems.
The derivation is given in Sec.\ref{sec_formula}.
We apply the formula to the paramagnetic state and the magnetic state
of LaFeAsO (Sec.\ref{sec_parent}) and 
the paramagnetic state of LaFeAsO$_{1-x}$F$_x$ (Sec.\ref{sec_doped}).
The electronic band structure of the parent compound
is determined by first principles calculations.
For doped systems, we use the five band model described in ref.\citenum{Kuroki08}.
We consider LaFeAsO$_{1-x}$F$_x$ because two-dimensionality is stronger
than other FeAs-based superconductors, such as BaFe$_2$As$_2$.
%Our formula is suitable for systems with strong two-dimensionality.
%We discuss the effect of Fermi surface topology on the interlayer magnetoresistance
%with in-plane magnetic fields.
%In sec.\ref{sec_formula} we derive the interlayer conductivity 
%for multi-band systems.
%The results for LaFeAsO is given in sec.\ref{sec_result}.
%Section \ref{sec_discussion} is devoted to discussion.

\section{Interlayer Conductivity Formula}
\label{sec_formula}
In this section, we derive the angular magnetoresistance formula
for multi-band systems.
We assume that the system has a structure of stacking conduction layers
in the $z$ axis.
The magnetic field ${\bf B} = (B_x, B_y, 0) = B(\cos \phi, \sin \phi,0)$ 
is assumed to be in the plane.
The Fermi surface topology is associated with the azimuthal angle $\phi$ dependence
of the interlayer magnetoresistance as we shall see below.
The vector potential ${\bf A}$ is taken as
${\bf{A}} = \left( B_y z, - B_x z, 0 \right)$.

The interlayer tunneling between $j$-th plane and $j+1$-th plane
is described by
\be
H_t  =  - t_c \sum\limits_{j,\nu,\sigma=\uparrow,\downarrow } {\int {d^2 {\bf{r}}} } 
\psi _{j + 1,\sigma}^{\left( \nu  \right)\dag } \left( {\bf{r}} \right)
\psi _{j,\sigma}^{\left( \nu  \right)} \left( {\bf{r}} \right)
\exp \left( {i\chi \left( {\bf{r}} \right)} \right) + h.c.,
\ee
where $\nu$ represents the band index.
%%%%%%%%%%%%%%%%%%%%%%%%%%%%%%%%%%%%%%%%%%%%%%%%%%%%%%%%%%%%%%%%%%%%%%%%%%%
% Revise 0728/09
%%%%%%%%%%%%%%%%%%%%%%%%%%%%%%%%%%%%%%%%%%%%%%%%%%%%%%%%%%%%%%%%%%%%%%%%%%%
%We include the in-plane magnetic field via a gauge transformation
%in which the wave function is multiplied by 
%$\exp \left( { - \frac{{ie}}{{c\hbar }}B_y zx 
%+ \frac{{ie}}{{c\hbar }}B_x zy} \right)$.
%The phase $\chi \left( {\bf{r}} \right)$ is associated with
%this gauge transformation:
%\be
%\chi \left( {\bf{r}} \right) = \frac{e}{{c\hbar }}a_c 
%\left( {B_y x - B_x y} \right).
%\ee
%with $a_c$ the interlayer spacing along the $z$ axis.
%%%%%%%%%%%%%%%%%%%%%%%%%%%%%%%%%%%%%%%%%%%%%%%%%%%%%%%%%%%%%%%%%%%%%%%%%%%
%{\bf 
Here we have included the effect of the in-plane magnetic field 
via a Pierls procedure:
\be
\chi \left( {\bf{r}} \right) = \frac{e}{{c\hbar }}a_c 
\left( {B_y x - B_x y} \right),
\ee
with $a_c$ the interlayer spacing along the $z$ axis.
%}
%%%%%%%%%%%%%%%%%%%%%%%%%%%%%%%%%%%%%%%%%%%%%%%%%%%%%%%%%%%%%%%%%%%%%%%%%%%
% End
%%%%%%%%%%%%%%%%%%%%%%%%%%%%%%%%%%%%%%%%%%%%%%%%%%%%%%%%%%%%%%%%%%%%%%%%%%%
%%%%%%%%%%%%%%%%%%%%%%%%%%%%%%%%%%%%%%%%%%%%%%%%%%%%%%%%%%%%%%%%%%%%%%%%%%%
% Revise 0728/09
%%%%%%%%%%%%%%%%%%%%%%%%%%%%%%%%%%%%%%%%%%%%%%%%%%%%%%%%%%%%%%%%%%%%%%%%%%%
%Using the $z$ axis current operator derived from $H_t$,
%the interlayer conductivity is calculated by the Kubo formula\cite{Mahan}
%{\bf 
The interlayer conductivity is calculated by the Kubo formula\cite{Mahan_kubo_formula},
%}
\be
\sigma _{zz}  = \mathop {\lim }\limits_{\omega  \to 0} \frac{{{\mathop{\rm Im}\nolimits} 
K_{zz}^R \left( \omega  \right)}}{\omega },
\ee
%{\bf 
where $K_{zz}^R \left( \omega  \right)$
is the Fourier transform of the following retarded current-current correlation function,
%}
\be
K_{zz}^R \left( {t - t'} \right) =  - \frac{i}{{\hbar S}}\theta 
\left( {t - t'} \right)\left\langle {T_\tau  
\left[ {J_z \left( t \right),J_z \left( {t'} \right)} \right]} \right\rangle, 
\ee
%{\bf 
with $S$ the area of the system.
We calculate $K_{zz}^R (t)$ in the Matsubara formalism. 
The correlation function in the Matsubara formalism is given by
%}
\be
K_{zz}^M \left( \tau  \right) = \frac{1}{S}\left\langle {T_\tau  J_z 
\left( \tau  \right)J_z \left( 0 \right)} \right\rangle 
\label{eq_KM}
\ee
%{\bf
Here the current operator is
%}
\be
J_z  = \frac{{iea_c t_c }}{\hbar }\sum\limits_{j,\nu ,\sigma } 
{\int {d^2 {\bf{r}}} \,\psi _{j + 1,\sigma }^{\left( \nu  \right)\dag } 
\left( {\bf{r}} \right)\,} \psi _{j,\sigma }^{\left( \nu  \right)} 
\left( {\bf{r}} \right)\,\exp \left( {i\chi \left( {\bf{r}} \right)} \right) + h.c.
\ee
%{\bf 
The function $K_{zz}^R \left( \omega  \right)$ is obtained by 
$K_{zz}^R \left( \omega  \right) = K_{zz}^M \left( {i\omega _n  \to \omega  + i\delta } \right)$
where $K_{zz}^M \left( {i\omega _n } \right) = \int_0^\beta  {d\tau } 
\exp \left( i\omega _n \tau \right)  K_{zz}^M \left( \tau  \right)$
and $\delta$ is a positive infinitesimal number.
For the non-interacting case, we obtain
%}
\bea
\sigma _{zz}  &=& \frac{{2\pi \hbar }}{S}\sum\limits_{{\bf{k}},{\bf{k'}},\nu,j,\sigma } 
{\int_{ - \infty }^\infty  {dE} \left( { - \frac{{\partial f}}{{\partial E}}} \right)\delta 
\left( {E - E_{{\bf{k}},\sigma }^{\left( \nu  \right)} } \right)\delta 
\left( {E_{{\bf{k}},\sigma }^{\left( \nu  \right)}  - E_{{\bf{k'}},\sigma }^{\left( \nu  \right)} } 
\right)} \nonumber \\ 
& &  \times \left| {\left\langle {{\bf{k'}},j + 1,\sigma } 
\right|J_z \left| {{\bf{k}},j,\sigma } \right\rangle } \right|^2.
\label{eq_szz_pre}
\eea
%{\bf 
where the $\nu$-th energy band is denoted by $E_{\bf k}^{(\nu)}$.
Substituting the matrix element of the current operaotr $J_z$ for the electron hopping 
from the $j$-th layer to the $j+1$-th layer,
%}
\be
\left\langle {{\bf{k'}},j + 1,\sigma } \right|J_z \left| {{\bf{k}},j,\sigma } \right\rangle  
= \frac{{iea_c t_c }}{\hbar }
\delta_{{\bf k},{\bf k}'-{\bf b}},
%\delta _{k'_x ,k_x -b_x} 
%\delta _{k'_y ,k_y  - b_y },
\ee
with ${\bf b}=ea_c(-B_y,B_x)/(c\hbar)$,
%{\bf
we obtain
%}
\bea
 \sigma _{zz}  &=&  \frac{{2e^2 t_c^2 a_c^2 }}{\hbar }N_z \sum\limits_{\nu ,\sigma } 
{\int {\frac{{d^2 {\bf{k}}}}{{\left( {2\pi } \right)^2 }}} \int_{ - \infty }^\infty  {dE} 
\left( { - \frac{{\partial f}}{{\partial E}}} \right)}  \nonumber \\ 
& &  \times \delta \left( {E - E_{{\bf{k}},\sigma }^{\left( \nu  \right)} } \right)
% Lorentz factor
\frac{\Gamma}{\left( E_{{\bf{k}},\sigma }^{\left( \nu  \right)}  
- E_{{\bf{k}} - {\bf{b}},\sigma }^{\left( \nu  \right)} \right)^2  + \Gamma^2 }.
\label{eq_sigma_zz_full}
\eea
%{\bf
Here we have replaced the second Dirac delta function in eq.\refeq{eq_szz_pre}
by a Lorentz function with $\Gamma =\hbar/(2\tau)$ 
to include the scattering effect upon hopping between layers.
%}
%%%%%%%%%%%%%%%%%%%%%%%%%%%%%%%%%%%%%%%%%%%%%%%%%%%%%%%%%%%%%%%%%%%%%%%%%%%
% End
%%%%%%%%%%%%%%%%%%%%%%%%%%%%%%%%%%%%%%%%%%%%%%%%%%%%%%%%%%%%%%%%%%%%%%%%%%%
Although it is possible to take different $\Gamma$'s 
for each band,
we take the same $\Gamma$ for all bands for simplicity.
%%%%%%%%%%%%%%%%%%%%%%%%%%%%%%%%%%%%%%%%%%%%%%%%%%%%%%%%%%%%%%%%%%%%%%%%%%%
% Revise 0728/09
%%%%%%%%%%%%%%%%%%%%%%%%%%%%%%%%%%%%%%%%%%%%%%%%%%%%%%%%%%%%%%%%%%%%%%%%%%%
%{\bf 
The formula (\ref{eq_sigma_zz_full}) is understood in a following way.
According to the semiclassical equations under a magnetic field,
an electron moves in ${\bf k}$-space upon hopping
between one layer to the adjacent layers.
The contribution to $\sigma_{zz}$ is large 
if the motion in ${\bf k}$-space is along the Fermi surface
because the Lorentz factor in eq.\refeq{eq_sigma_zz_full}
takes the maximum value.
Since the direction of the motion is perpendicular to the 
in-plane magnetic field, dominant contributions arise 
if the magnetic field is perpendicular to the flat regions
of the Fermi surface.
%}

At zero temperature, the integration over the Brillouin zone 
is transformed to the integration
along the Fermi surfaces as follows
\bea
\sigma _{zz}  
& = & \frac{{e^2 }}{{2\pi }}
\left( {\frac{{t_c a_c }}{\hbar }} \right)^2 N_z \sum\limits_\nu  
%\oint\limits_{E_{\bf{k}}^{(\nu )}  = E_F } {d\ell _{\bf{k}} } 
\int\limits_{E_{\bf{k}}^{(\nu )}  = E_F } {d\ell _{\bf{k}} } 
\frac{1}{{\left| {{\bf v}_{\bf{k}}^{\left( \nu  \right)} } \right|}}
\nonumber \\
& & \times \frac{{\Gamma /\pi }}{\left( 
%{\frac{{e a_c }}{{c\hbar }} 
%\frac{{\partial E_{\bf{k}}^{\left( \nu  \right)} }}
%{{\partial {\bf{k}}}} 
\frac{e}{c} a_c {\bf v}_{\bf k}^{(\nu)} \times {\bf{B}} \right)^2  
+ \Gamma ^2},
\label{eq_sigma_zz}
\eea
where ${\bf v}_{\bf k}^{(\nu)} = (1/\hbar)\partial E_{\bf k}^{(\nu )}/\partial {\bf k}$.
From the form of the Lorentz factor, it is clear that
dominant contributions to $\sigma_{zz}$ arise
from the Fermi surface parts with the Fermi velocity 
being parallel to the magnetic field,
which is consistent with the intuitive picture mentioned above.
The weight of the contribution is determined 
by the factor $1/|{\bf v}_{\bf k}^{(\nu )}|$ as in the density of states.
Anisotropy in $\sigma_{zz}$ detects 
deviation of the Fermi surface shape from the circular shape.
If a Fermi surface is completely circular,
such a Fermi surface contributes to $\sigma_{zz}$
only as the constant value independent of the orientation of 
the magnetic field.
In order to calculate $\sigma_{zz}$ from eq.(\ref{eq_sigma_zz})
we need values of ${\bf v}_{\bf k}^{(\nu)}$
%$\partial E_{\bf k}^{(\nu )}/\partial {\bf k}$,
and the Fermi surface data for each band.
For Fermi surfaces which are not given by
separated lines, one needs to calculate 
eq.(\ref{eq_sigma_zz_full}).

Note that we have neglected the Zeeman energy
contribution in eq.(\ref{eq_sigma_zz}).
To be precise, we need the energy contour
at $E^{(\pm )}=E_F \pm g\mu_B B$.
However, the Zeeman energy is much less than
the Fermi energy.
Therefore, we may neglect the Fermi surface change
associated with the Fermi energy shift
by the Zeeman energy.

\section{Interlayer Conductivity of LaFeAsO}
\label{sec_parent}
\begin{figure}[htbp]
  \begin{center}
  %\begin{tabular}{cc}
  \subfigure[]{
    \includegraphics*[height=5cm,keepaspectratio]{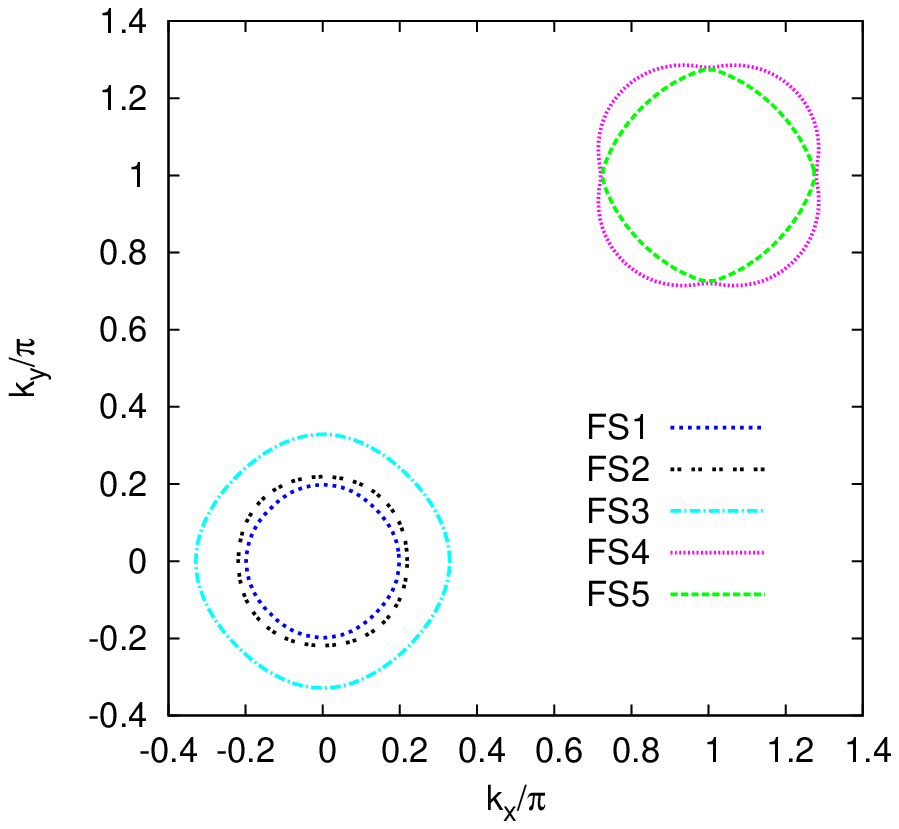}
  }
  \subfigure[]{
    \includegraphics*[height=5cm,keepaspectratio]{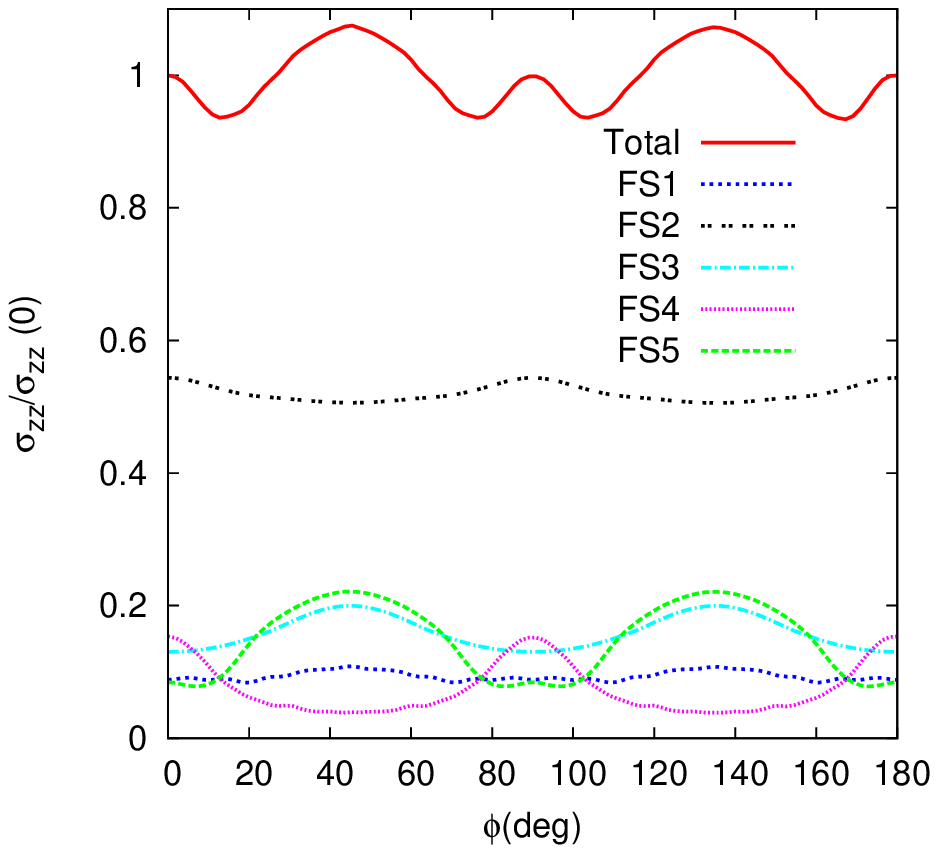}
  }
%\end{tabular}
%\caption{}
\end{center}
\caption{ 
(Color online)
The paramagnetic state Fermi surface (a) and the normalized interlayer conductivity(b).
}
\label{figNormal}
\end{figure}

For the application to LaFeAsO, we compute the electronic band 
structure by first principles calculations.
The Fermi surfaces were obtained by the first-principles density 
functional calculation package VASP.\cite{Kresse93,*Kresse96,*Kresse96b}
%\subsection{Normal state}
Figure \ref{figNormal}(a) shows the Fermi surface of the paramagnetic state
and Fig.\ref{figNormal}(b) shows the interlayer conductivity 
for $\Gamma=2\times 10^{-3}$
and $B=10$T which corresponds to $\omega_c \tau=eB\tau/(mc)=0.29$.
(Hereafter we measure $\Gamma$ in units of ${\rm eV}$.)
%functional calculation package VASP.\cite{VASP}
Each band contribution to the interlayer conductivity is also shown.
The values are normalized by the total interlayer conductivity
at $\phi=0^\circ$.
The broad peaks in the interlayer conductivity at $\phi  = 45^\circ$
and $\phi  = 135^\circ$ are associated with the contributions from
the Fermi surfaces FS3 and FS5 because 
those Fermi surfaces have flat regions
whose Fermi velocities are in these directions.
The peak at $\phi=90^\circ$ is associated with the flat region
in the Fermi surface FS4.
The Fermi surfaces FS1 and FS2 are nearly circular.
Therefore, their interlayer conductivities are almost featureless.
The largest contribution to $\sigma_{zz}$ comes from FS2
but other band features are clearly seen because
anisotropy in the FS2 contribution is small.

%=====================================================================
% Normal state FS and MR (two figures)
%=====================================================================
% code: rd2_3.cc
% gnuplot file: gpf_rd2_3c.plt
% B=10, Gamma = 2.0e-3
%%%%%%%%%%%%%%%%%%%%%%%%%
%=====================================================================

\begin{figure}[htbp]
  \begin{center}
  %\begin{tabular}{cc}
  \subfigure[]{
    \includegraphics*[height=5cm,keepaspectratio]{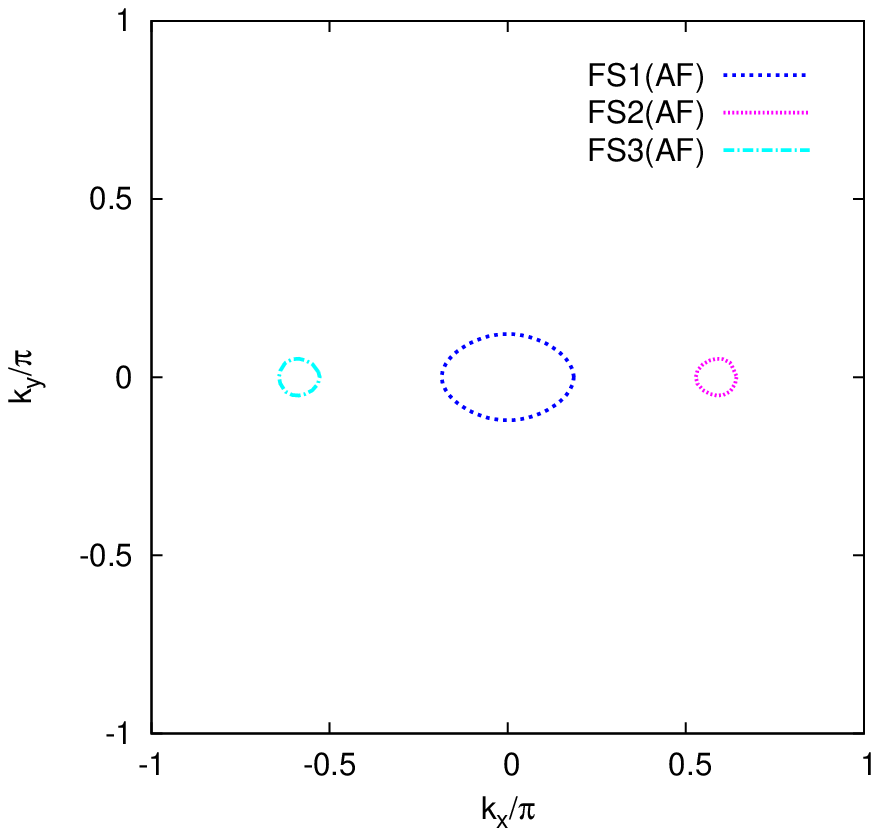}
  }
  \subfigure[]{
    \includegraphics*[height=5cm,keepaspectratio]{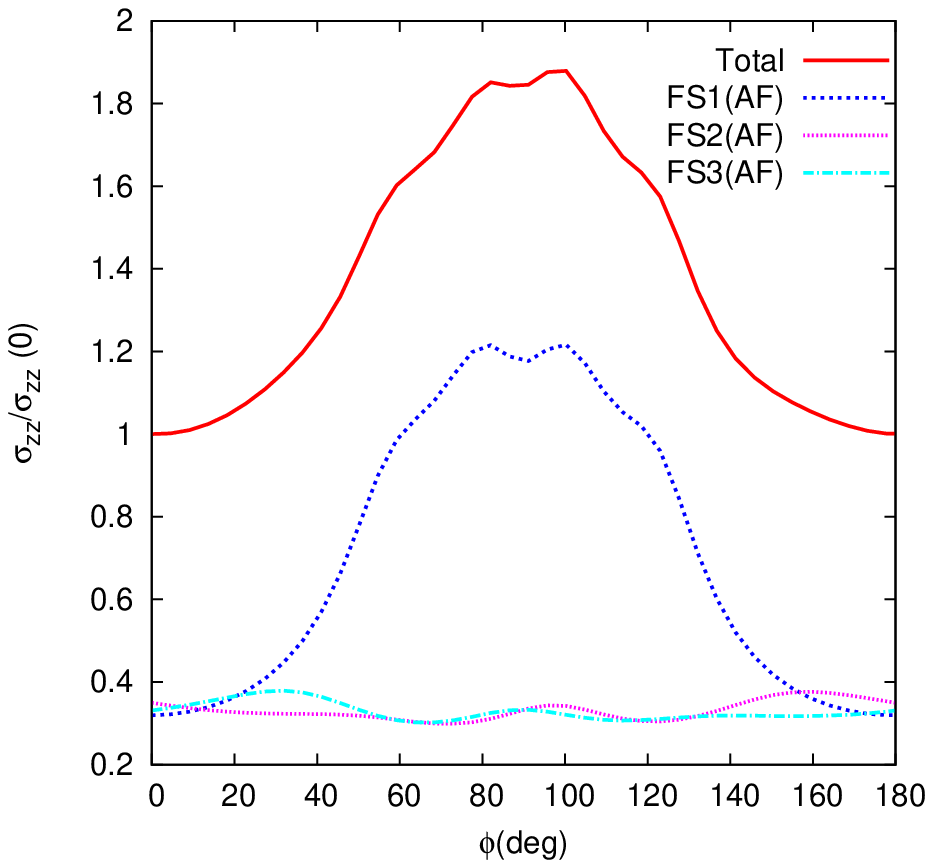}
  }
%\end{tabular}
%\caption{}
\end{center}
\caption{ 
(Color online)
The Fermi surface of the antiferromagnetic state (a) and 
the normalized interlayer conductivity(b).
}
\label{figAF}
\end{figure}

%\subsection{AF state}
The Fermi surface and the interlayer conductivity
in the magnetically ordered state is shown in
Figs.\ref{figAF}(a) and \ref{figAF}(b).
The magnetic ordered state is a collinear 
antiferromagnetic order.\cite{Ishibashi08}
The interlayer conductivity is dominated by the contribution
from the Fermi surface around the $\Gamma$ point (FS1).
The broad peak at $\phi=90^\circ$ is associated with the flat 
regions in FS1.
The small Fermi surfaces FS2 and FS3 are almost featureless
because they are nearly circular.
%=====================================================================
% AF state FS and MR (two figures)
%=====================================================================
% script: unif1_1.pl, gpf_rd4_0b.plt
% Gamma=2.0e-3
% 
%=====================================================================

%\subsection{Difference}
A clear difference between the paramagnetic state
and the antiferromagnetic state is the oscillation period 
difference in $\sigma_{zz}$.
In the paramagnetic state, the oscillation period is $90^\circ$.
By contrast the oscillation period of the antiferromagnetic state
is $180^\circ$.
Therefore, the presence of the magnetic ordering is 
distinguished from the paramagnetic state
by measuring the oscillation period of $\sigma_{zz}$.

\begin{figure}[htbp]
  \begin{center}
  %\begin{tabular}{cc}
  \subfigure[]{
    \includegraphics*[height=5cm,keepaspectratio]{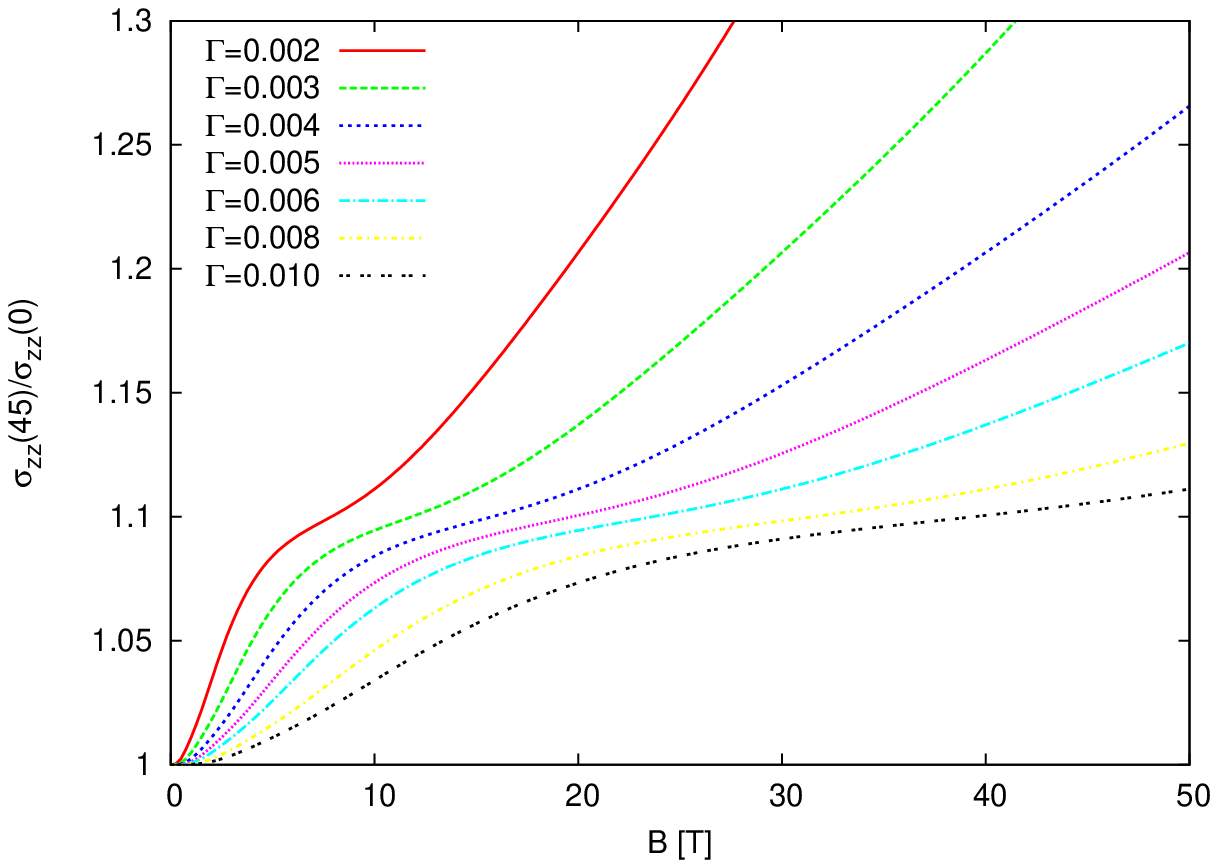}
  }
  \subfigure[]{
    \includegraphics*[height=5cm,keepaspectratio]{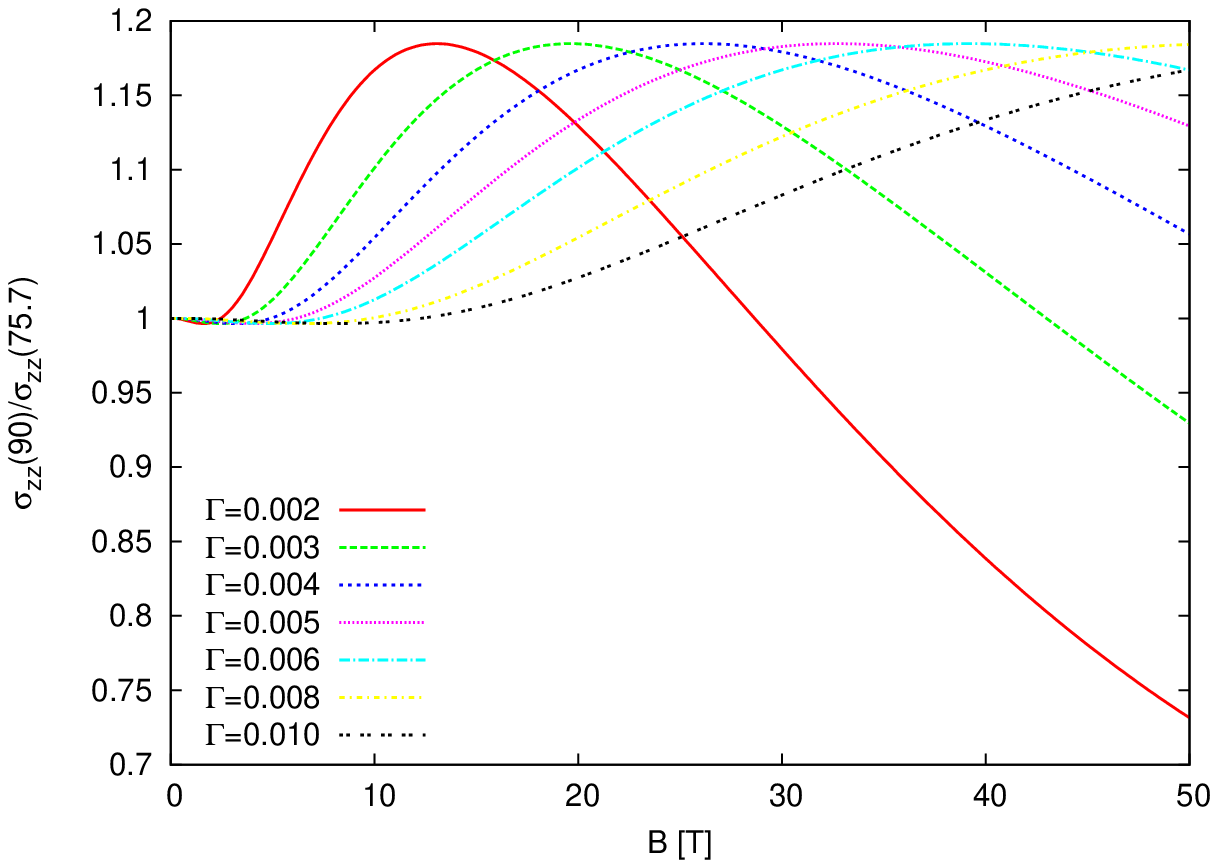}
  }
%\end{tabular}
%\caption{}
\end{center}
\caption{ 
(Color online)
Magnetic field dependence of $\sigma_{zz}(45^\circ)/\sigma_{zz}(0^\circ)$(a)
and $\sigma_{zz}(90^\circ)/\sigma_{zz}(75.7^\circ)$(b)
in the paramagnetic state.
}
\label{fig_normal_B}
\end{figure}

\begin{figure}[htbp]
  \begin{center}	
   \includegraphics*[height=5cm,keepaspectratio]{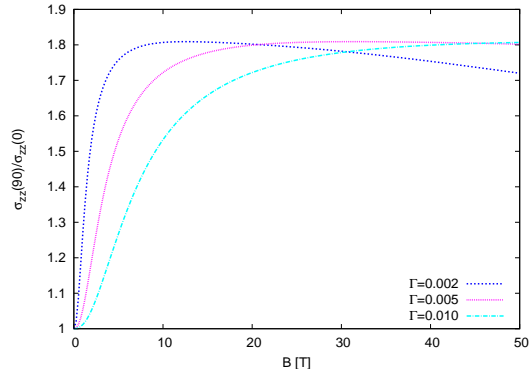}
  \end{center}
\caption{ 
(Color online)
Magnetic field dependence of $\sigma_{zz}(90^\circ)/\sigma_{zz}(0^\circ)$
in the magnetic state.
}
\label{fig_AF_B}
\end{figure}

%\subsection{B dependence}
In order to make clear the condition of observing the oscillations
in the interlayer conductivity,
we compute the magnetic field dependence of the peak height
for several values of $\Gamma$.
The onset of the anisotropy depends on $\Gamma$.
Therefore, the measurements of the magnetic field dependence
of the anisotropy provides information about the scattering.
Figure \ref{fig_normal_B}(a) shows the magnetic field
dependence of the ratio of the peak in $\sigma_{zz}$ at $45^\circ$
and the value at $0^\circ$ in the paramagnetic state.
The ratio increases monotonically as we increase 
the magnetic field.
Figure \ref{fig_normal_B}(b) shows the magnetic field 
dependence of the ratio of the peak at $90^\circ$
and the dip at $75.7^\circ$.
Since the magnetic field effect on those angles are not simple,
the ratio shows a non-monotonic behavior.
In principle one can deduce $\Gamma$ by measuring 
these two ratios by comparing the theoretical prediction
based on first principles calculations with the experiments.
Figure \ref{fig_AF_B} shows the magnetic field dependence
of the ratio of the peak in $\sigma_{zz}$ at $90^\circ$ and 
the value at $0^\circ$ in the magnetically ordered state.
%=====================================================================
% B-dep. (Normal)
%=====================================================================
% Gamma=5.0e-3
% 45-0,  90-75.7(0.42pi)
%=====================================================================

%=====================================================================
% B-dep. (AF)
%=====================================================================
% Gamma=5.0e-3, 1.0e-2, 1.5e-2
% 90-0
%=====================================================================

\section{Doped Systems}
\label{sec_doped}
\begin{figure}[htbp]
  \begin{center}
  \begin{tabular}{cc}
  \begin{minipage}{0.5\hsize}
  \subfigure[$n=5.9$]{
    \includegraphics*[height=5cm,keepaspectratio]{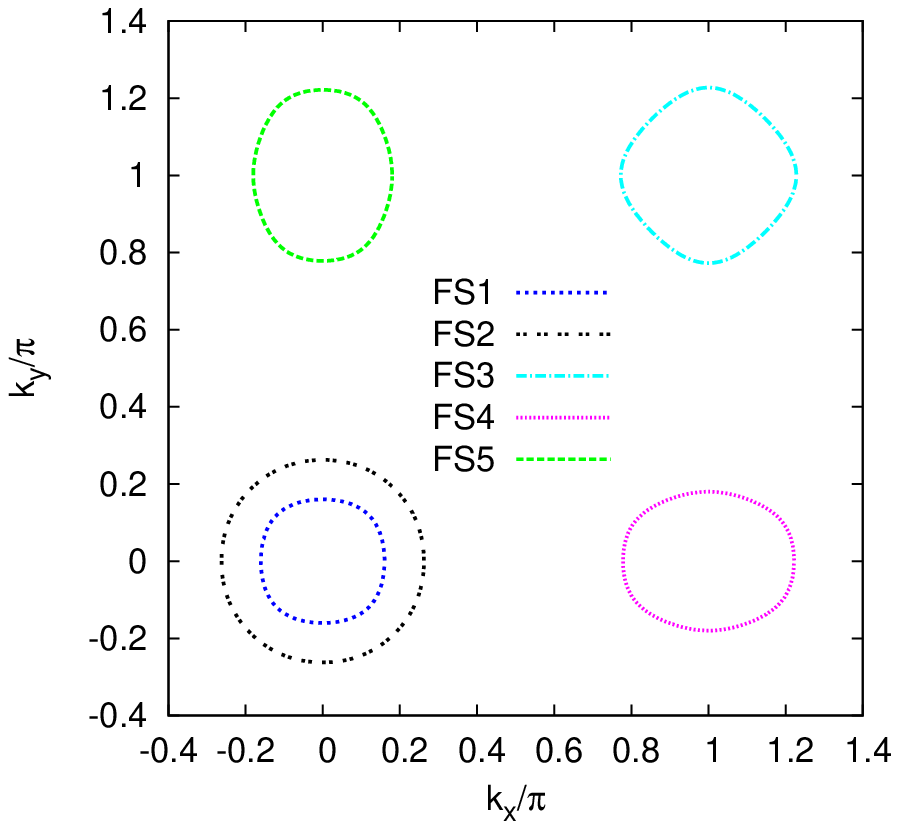}
  }
  \end{minipage}
  \begin{minipage}{0.5\hsize}
  \subfigure[$n=6$]{
    \includegraphics*[height=5cm,keepaspectratio]{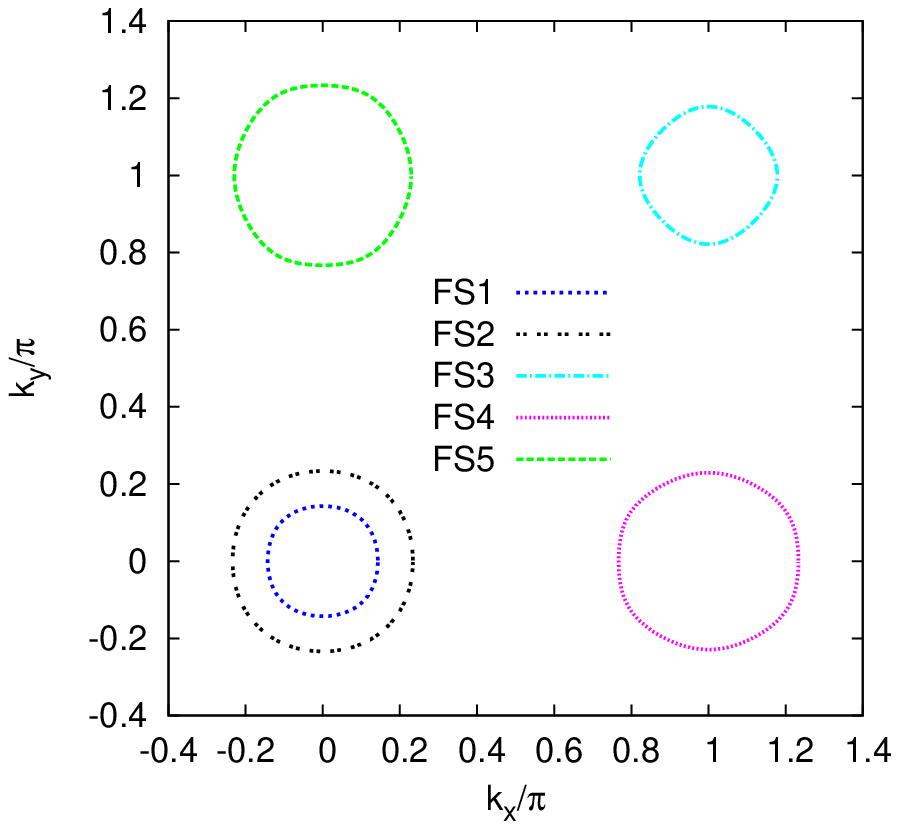}
  }
  \end{minipage}
  \\
  %%%%%%%%%%%%%%%%%%%%%%%%%%%%%%%%%%%%%%%%%%%%%%%%%%%%
  \begin{minipage}{0.5\hsize}
  \subfigure[$n=6.2$]{
    \includegraphics*[height=5cm,keepaspectratio]{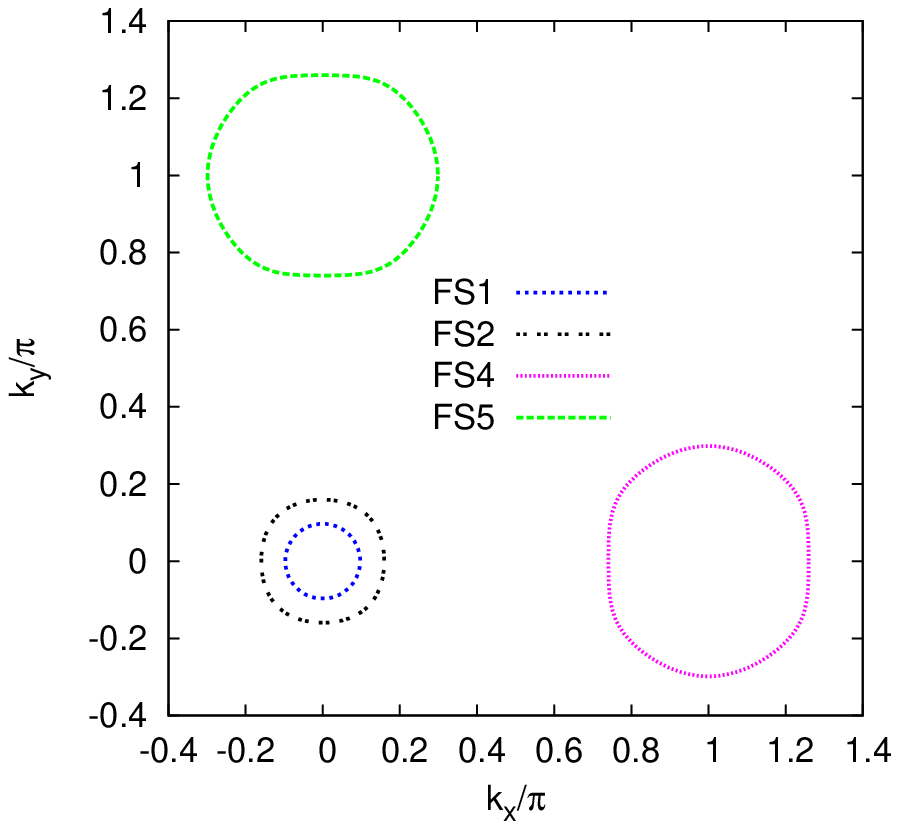}
  }
  \end{minipage}
  \begin{minipage}{0.5\hsize}
  \subfigure[$n=6.4$]{
    \includegraphics*[height=5cm,keepaspectratio]{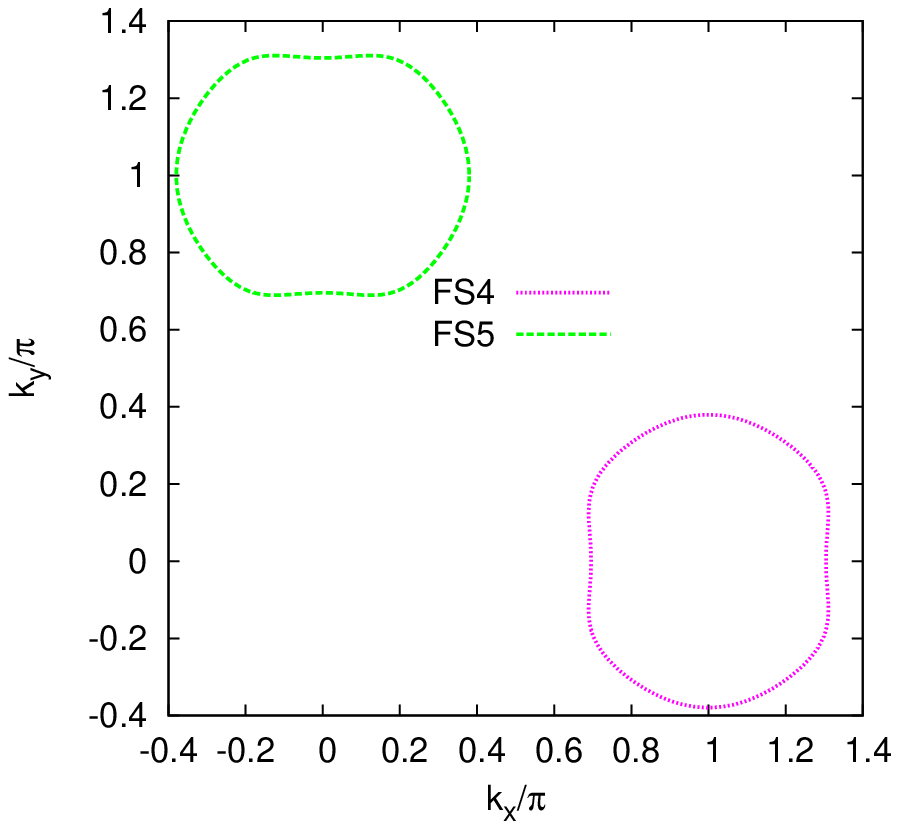}
  }
  \end{minipage}
\end{tabular}
\end{center}
\caption{ 
(Color online)
Electron density dependence of the Fermi surface calculated from
the five-band model by Kuroki {\it et al}.\cite{Kuroki08}
}
\label{fig_fs_change}
\end{figure}

\begin{figure}[htbp]
  \begin{center}
  \begin{tabular}{cc}
  \begin{minipage}{0.5\hsize}
  \subfigure[$n=5.9$]{
    \includegraphics*[height=5cm,keepaspectratio]{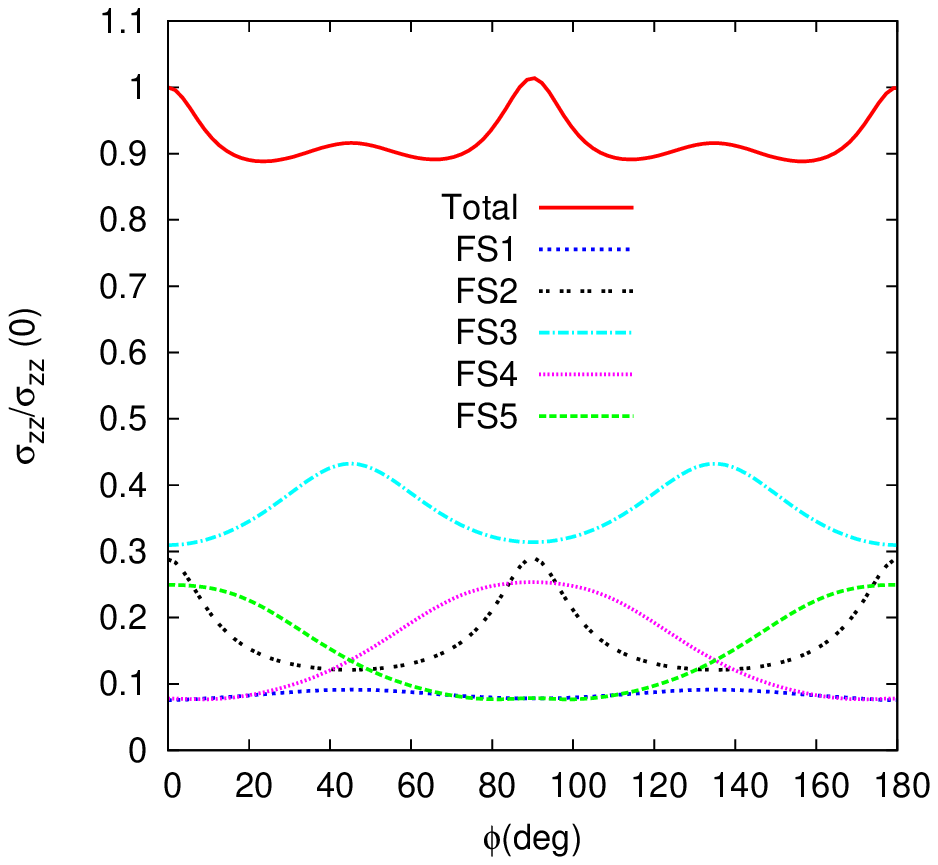}
  }
  \end{minipage}
  \begin{minipage}{0.5\hsize}
  \subfigure[$n=6$]{
    \includegraphics*[height=5cm,keepaspectratio]{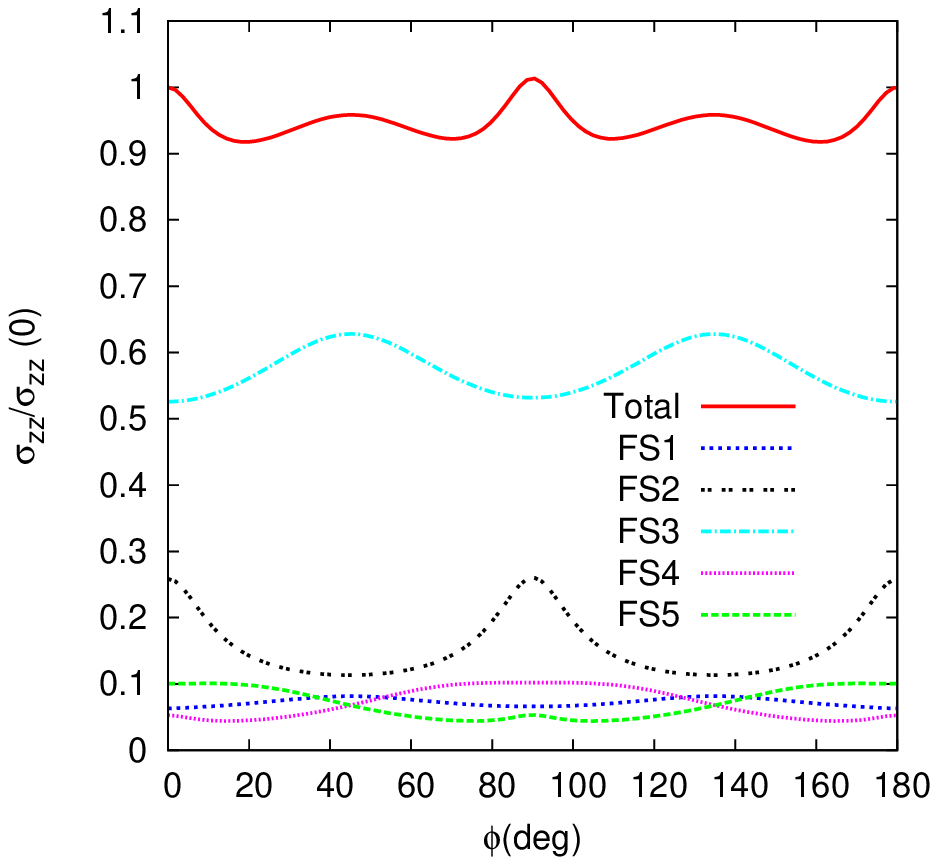}
  }
  \end{minipage}
  \\
  %%%%%%%%%%%%%%%%%%%%%%%%%%%%%%%%%%%%%%%%%%%%%%%%%%%%
  \begin{minipage}{0.5\hsize}
  \subfigure[$n=6.2$]{
    \includegraphics*[height=5cm,keepaspectratio]{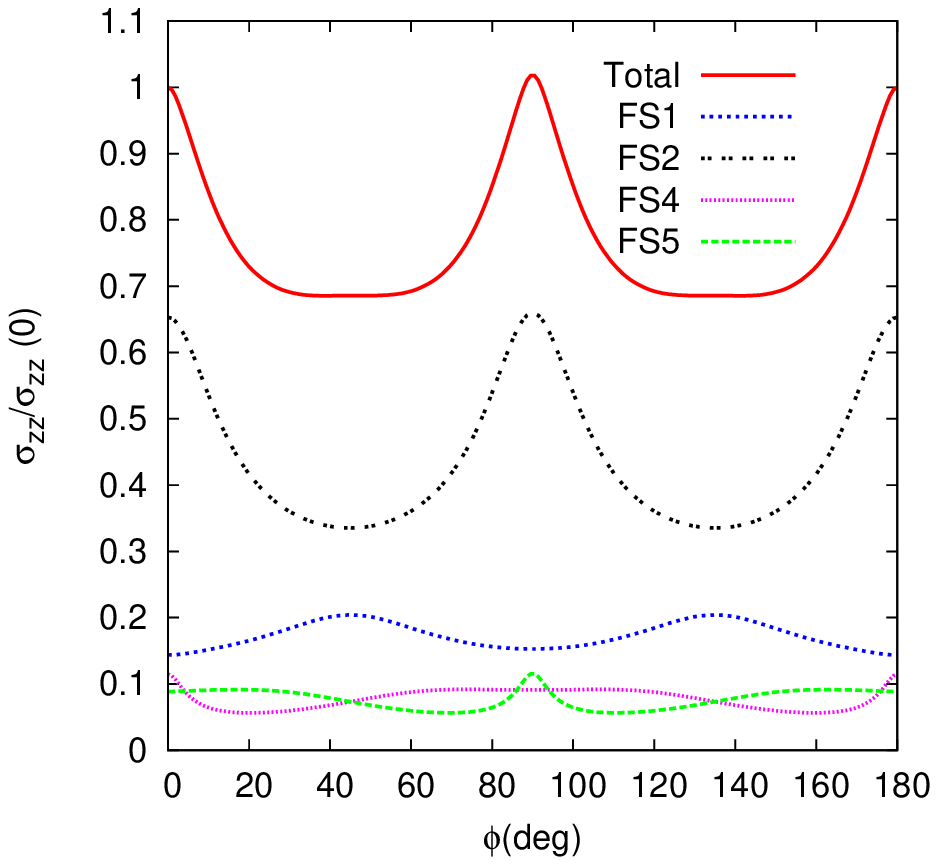}
  }
  \end{minipage}
  \begin{minipage}{0.5\hsize}
  \subfigure[$n=6.4$]{
    \includegraphics*[height=5cm,keepaspectratio]{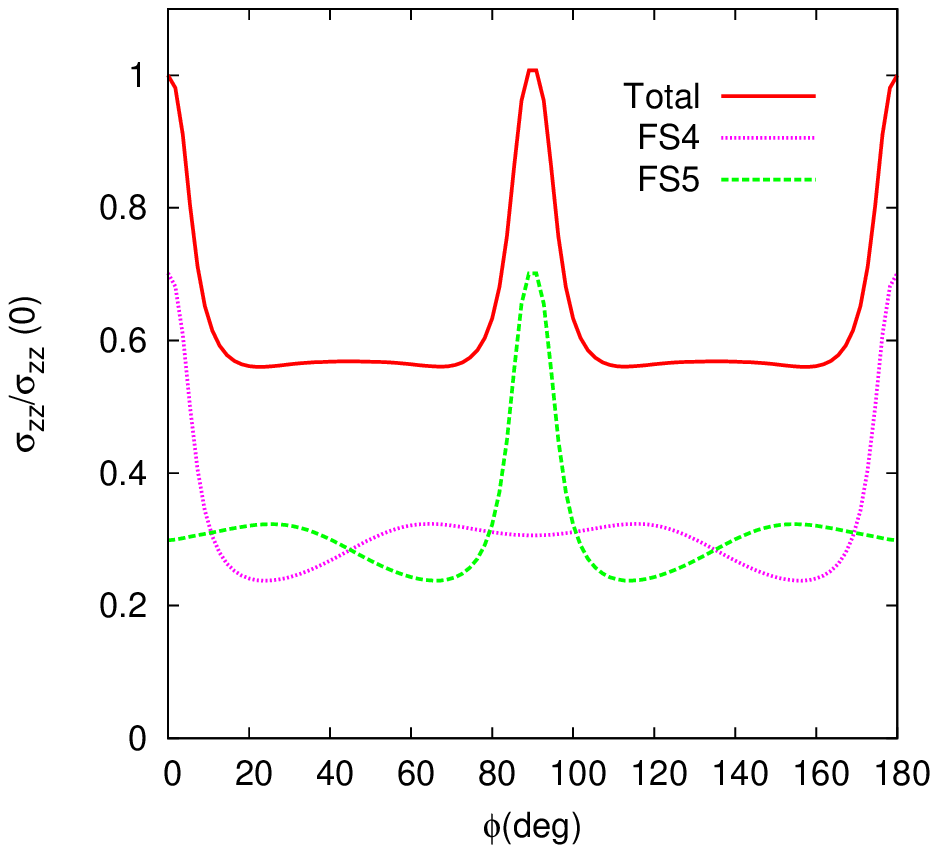}
  }
  \end{minipage}
\end{tabular}
\end{center}
\caption{ 
(Color online)
Electron density dependence of the Fermi surface calculated from
the five-band model by Kuroki {\it et al}.\cite{Kuroki08}
}
\label{fig_mr_change}
\end{figure}

In this section we consider the doping dependence of the interlayer
conductivity.
Our consideration is based on the rigid band picture, and we take 
the five-band model Hamiltonian from Kuroki {\it et al}.\cite{Kuroki08}
Figure \ref{fig_fs_change} shows the electron density dependence 
of the Fermi surface and Fig.\ref{fig_mr_change} shows 
the interlayer conductivity for each electron concentration
with $B=10{\rm T}$ and $\Gamma=2 \times 10^{-3}$.
Note that the azimuthal angle $\phi$ in Fig.\ref{fig_mr_change}
is shifted by $45^\circ$ compared to Fig.\ref{figNormal} because 
of the difference of the unit cells.
(In our first principles calculations, 
the unit cell contains two Fe atoms and two As atoms.
Meanwhile in the five band model of ref.\citenum{Kuroki08},
the unit cell contains one Fe atom.
The Fe-Fe bond direction differs by $45^\circ$ between
the two unit cells.)
If we compare Fig.\ref{fig_mr_change}(b) with Fig.\ref{figNormal}(b),
they are qualitatively the same.
The peaks at $\phi=45^\circ$ and $\phi=135^\circ$ in Fig.\ref{figNormal}(b)
are observed at $\phi=0^\circ$, $90^\circ$, and $180^\circ$ 
in Fig.\ref{fig_mr_change}(b).
The peak at $\phi=90^\circ$ in Fig.\ref{figNormal}(b) is observed
at $\phi=45^\circ$ and $\phi=135^\circ$
in Fig.\ref{fig_mr_change}(b).
However, the peak widths are different.
The discrepancy is mainly associated with the Fermi surface shape difference
between FS4 and FS5 in Fig.\ref{figNormal}(a) and FS4 and FS5 
in Fig.\ref{fig_fs_change}(b).
In addition there are non-negligible discrepancies in the Fermi velocities 
between the five band model and the first principles calculation.
Those discrepancies lead to the differences of each band's weight
in the interlayer conductivity.
The Fermi velocities are the derivatives of the dispersion energies
with respect to $k_x$ or $k_y$.
However, we may not expect that the five band model should reproduce
derivatives of the energy dispersions of the first principles calculation.
Although there is some quantitative differences, the five band model
reproduces the first principles result qualitatively.

In hole doped case, the result of the interlayer conductivity 
is similar to that of $n=6$:
For 10$\%$ hole doping case ($n=5.9$)
the Fermi surface Fig.\ref{fig_fs_change}(a) is similar to
$n=6$ case (Fig.\ref{fig_fs_change}(b))
and so their interlayer conductivities are similar 
(Figs.\ref{fig_mr_change}(a) and \ref{fig_mr_change}(b)).
Upon further hole doping, the electron Fermi surfaces FS4 and FS5 shrink.
But the shrink of those Fermi surfaces does not lead to 
qualitative difference of the interlayer conductivity.

In the electron doped side, for $n=6.2$, the hole Fermi surface around 
$(\pi, \pi)$ (FS3) disappears (Fig.\ref{fig_fs_change}(c)).
The effect of this disappearance of the Fermi surface FS3 is clearly seen
in the interlayer conductivity Fig.\ref{fig_mr_change}(c):
The peak at $\phi=45^\circ$ and $\phi=135^\circ$ 
in Figs.\ref{fig_mr_change}(a) and \ref{fig_mr_change}(b),
which is associated with the flat regions of the Fermi surface FS3,
vanishes in Fig.\ref{fig_mr_change}(c).
For $n=6.4$ the two hole Fermi surfaces around $(0,0)$ (FS1 and FS2)
disappear, and the peak at $\phi=90^\circ$ in the interlayer conductivity
becomes sharp (Fig.\ref{fig_mr_change}(d))
because of large flat regions of FS4 and FS5 (Fig.\ref{fig_fs_change}(d)).

%=====================================================================
% Fermi surface change
%=====================================================================
% code: rd5_6.cc
% Data: Data0525_09/ (created by fb4_2_2.cc)
%=====================================================================

%=====================================================================
% Interlayer conductivity
%=====================================================================
% code: rd5_6.cc -> rd5_6_corr.cc, gpf_rd5_6_corr.plt
% Data: Data0525_09/ (created by fb4_2_2.cc)
%=====================================================================

\section{Summary and Discussion}
In this paper, we have proposed a interlayer conductivity formula
under an in-plane magnetic field
to study the Fermi surface topology of layered multiband systems.
Applications of the formula to one of the FeAs-based superconductors,
LaFeAsO$_{1-x}$F$_x$, have been discussed.
Although the total interlayer conductivity 
is given by the sum of all band contributions,
it is possible to extract information about
the Fermi surface topology from the peak
positions in the interlayer conductivity
as a function of the azimuthal angle of the magnetic field.
Flat regions of the Fermi surface give rise
to peaks in the interlayer conductivity,
while nearly circular Fermi surfaces do not
show oscillating patterns.
We have discussed the difference between the paramagnetic
state and the magnetic state of the parent compound LaFeAsO.
As for the presence of the magnetic state,
the oscillation period is $180^\circ$
for the magnetic state but $90^\circ$
for paramagnetic states.
To study the doping effect, we have taken the five band model
of ref.\citenum{Kuroki08} and discussed the Fermi surface 
topology change upon doping and how such changes are 
reflected to the interlayer conductivity.

%\subsection{$t_c/t$ effect}
In the analysis above, the interlayer coupling has been taken into account
through the interlayer current.
The energy dispersion along the $z$ axis is not included.
If the electron band dispersion along the $z$ axis is simply represented by
an additional term of $\epsilon_k^{(c)} = -2t_c \cos k_z$, 
%%%%%%%%%%%%%%%%%%%%%%%%%%%%%%%%%%%%%%%%%%%%%%%%%%%%%%%%%%%%%%%%%%%%%%%%%%%
% Revise 0729/09
%%%%%%%%%%%%%%%%%%%%%%%%%%%%%%%%%%%%%%%%%%%%%%%%%%%%%%%%%%%%%%%%%%%%%%%%%%%
%{\bf 
then the Lorentz function factor in eq.\refeq{eq_sigma_zz}
has dominant contributions at $k_z=0,\pi$.
However, if the Fermi energy is much larger than $t_c$,
the Fermi surface topology change is negligible.
For such a simple energy dispersion form in the $k_z$-direction,
one may apply the Shockley's tube-integral formula \cite{Shockley_tube_integral_formula} 
to analyze the angle-dependent magnetoresistance oscillations.
The angle-dependent magnetoresistance peaks
are connected with the vanishing of the electronic group velocity
perpendicular to the layers. \cite{Yamaji89}
Since the vanishing condition depends on the Fermi wave-vector,
which is azimuthal angle dependent, it is possible to extract
further information about the Fermi surface topology.\cite{Nam01}
%The analysis made in Sr$_2$RuO$_4$ showed that the results 
%were in good agreement with the quantum oscillation results.
%\cite{Bergemann03}
%}
%
%The three dimensionality affects the result if the $k_z$ dependence
%is significant.
%%%%%%%%%%%%%%%%%%%%%%%%%%%%%%%%%%%%%%%%%%%%%%%%%%%%%%%%%%%%%%%%%%%%%%%%%%%
% End
%%%%%%%%%%%%%%%%%%%%%%%%%%%%%%%%%%%%%%%%%%%%%%%%%%%%%%%%%%%%%%%%%%%%%%%%%%%
%In such a case, we need to include the full three dimensional 
%energy dispersion in the band energy, $E_{\bf k}^{(\nu)}$
%in eq.(\ref{eq_sigma_zz_full}).

%\subsection{scattering}
As for the scattering,
we have assumed the same scattering time for all bands for simplicity.
However, the scattering time could be different for each band.
In this sense, the measurement of the interlayer magnetoresistance 
is useful for the determination of the scattering time,
because oscillating patterns are discernible
from a certain magnetic field.
Although the onset is not directly related to $\Gamma$,
one can extract $\Gamma$ by comparing first principles calculation
results with the experiments.

%\subsection{Temperature effect}
The results above have been obtained at zero temperature.
In order to evaluate the finite temperature effect precisely,
we need to sum over contributions from energy contours
with including the first derivative of the Fermi-Dirac distribution function 
with respect to the energy.
The finite temperature effect can be estimated for
a simple energy dispersion case.
If the energy dispersion is well approximated by
$\epsilon_k = \hbar^2 (k_x^2 + \lambda k_y^2)/(2m)$,
then the amplitude of the conductivity is somewhat suppressed
for finite temperatures
but the characteristic temperature for the suppression
is scaled by the Fermi energy.
Furthermore, the temperature dependent factor is factorized out
from the magnetic field dependent part.
For energy dispersions with complicated $k_x$ and $k_y$ dependences,
this simple argument does not apply.
However, we expect that the finite temperature effect on $\Gamma$
is more significant than that on the other contributions in the 
interlayer conductivity.

%\subsection{Superconductivity}
The interlayer magnetoresistance measurement can be used
to find nodes in the superconducting state if there are nodes in the gap
as discussed by Bulaevskii {\it et al}.\cite{Bulaevskii99}
If the $z$ axis current exceeds the Josephson critical current,
then one can measure the interlayer magnetoresistance.
In the presence of the gap nodes, the interlayer magnetoresitance
detects the nodes if the applied magnetic field is parallel to
the normal state Fermi velocity at the node.
However, the formula given in ref.\citenum{Bulaevskii99}
is not directly applicable to the $s\pm$-wave gap
proposed for FeAs-based superconductivity.\cite{Mazin08,Kuroki08}

%advantages
Finally we comment on advantages of using the interlayer magnetoresistance
for investigating Fermi surface topology.
Although ARPES is a powerful measurement to extract information
about Fermi surface topology, 
the interlayer magnetoresistance experiments are applicable
to more compounds than ARPES,
since it is not necessary to cleave a sample to obtain a clean surface.
In addition, the interlayer magnetoresistance is associated with
the bulk properties.
Therefore, the measurement is not sensitive to the surface condition.
There are now many FeAs-based superconductors.
In order to make clear common underlying electronic properties
for the mechanism of superconductivity,
we need to compare two different systems.
In this sense, the interlayer magnetoresistance is useful 
because of its wide applicability.

%\section*{Acknowledgment}
We would like to thank T. Shibauchi for drawing our attention to
ref.\citenum{Bulaevskii99}.
This work was supported by the Grant-in-Aid for Scientific Research
from the Ministry of Education, Culture, Sports, Science and Technology (MEXT) 
of Japan, the Global COE Program 
"The Next Generation of Physics, Spun from Universality and Emergence," 
and Yukawa International Program for Quark-Hadron Sciences at YITP.

%\appendix
%%\section{Sample}

\bibliography{../../../references/tm_library2}

\end{document}